\documentclass[12pt]{iopart}

\usepackage{iopams}

\usepackage{amssymb}

\usepackage[english]{babel}
\usepackage[latin1]{inputenc}

\newcommand{\be}[1]{\begin{equation}\label{#1}}
\newcommand{\ee}{\end{equation}}
\newcommand{\ba}[1]{\begin{eqnarray}\label{#1}}
\newcommand{\ea}{\end{eqnarray}}
\newcommand{\rf}[1]{(\ref{#1})}
\newcommand{\nn}{\nonumber}

\renewcommand{\phi}{\varphi}

\def\p{\partial}
\def\a{\alpha}

\def\d{\delta}

\def\G{\Gamma}
\def\sg{\sigma}

\def\lb{\lambda}

\def\vfi{\varphi}

\def\ra{\rangle}
\def\la{\langle}

\def\sg{\sigma}

\newcommand{\wt}{\widetilde}

\def\wt{\widetilde}

\def\ci{\,\mbox{\rm Ci\,}}
\def\si{\,\mbox{\rm Si\,}}

\def\tfrac#1#2{{\textstyle {#1 \over #2}}}%
\newcommand{\bI}{\mathbf{I}}

\newcommand{\tb}[1]{\textbf{#1}}

\begin{document}

\normalfont

\title[Manifestation of spectral singularity]{Hermitian Hamiltonian
equivalent to a given non-Hermitian one.
Manifestation of spectral singularity
}

\author{\large Boris F. Samsonov}

\address{Physics Department, Tomsk State University, 36 Lenin Avenue,
634050 Tomsk, Russia}

\eads{\mailto{samsonov@phys.tsu.ru}}


\begin{abstract}
One of the simplest non-Hermitian Hamiltonians first proposed by
Schwartz (1960 {\it Commun. Pure Appl. Math.} \tb{13} 609)
which may possess a spectral singularity is analyzed from the
point of view of non-Hermitian generalization of quantum mechanics. It is
shown that $\eta$ operator, being a second order differential operator, has
supersymmetric structure. Asymptotic behavior of eigenfunctions of a
Hermitian Hamiltonian equivalent to the given non-Hermitian one is found.
As a result the corresponding scattering matrix and cross section
are given explicitly. It is demonstrated that the possible presence
of the spectral singularity in the spectrum of the non-Hermitian
Hamiltonian may be detected as a resonance in the scattering
cross section of its Hermitian counterpart. Nevertheless, just at the
singular point the equivalent Hermitian Hamiltonian becomes undetermined.
\end{abstract}

\section{Introduction}

Recently, one can notice a growing interest to non-Hermitian
Hamiltonians possessing
a real spectrum and
spectral singularities
\cite{collection}-\cite{my2005}.
Probably this is due to a remark that they may produce a resonance-like
effect in some optical experiments and may find an optical realization as a
certain type of lasing effect that occurs at the threshold gain
 \cite{Most2}.

This is well known that for any selfadjoint
(or essentially selfadjoint)
scattering Hamiltonian, continuous spectrum
eigenfunctions can be expressed in terms of Jost solution $f(k,x)$
 and Jost function $F(k)$, which, in the simplest case,
is the Jost solution taken at $x=0$,
$F(k)=f(k,0)$, see e. g. \cite{Levitan}.
One of the characteristic
features of any selfadjoint scattering Hamiltonian
is that its Jost function never
vanishes if $k$ is a spectral point, $F(k)\ne0$ $\forall k>0$
\cite{Levitan}.

An essential feature of the spectral singularity $k_0$ is that this
point belongs to
 a continuous part of the spectrum of a non-Hermitian Hamiltonian $H$
 and corresponding Jost function vanishes at this point $F(k_0)=0$.
As a result there is no way to construct a Hermitian%
\footnote{
An operator
$A$ in a Hilbert space is said to be selfadjoint
if $A=A^\dag$ where $A^\dag$ is Hermitian
adjoint to $A$. This definition assumes that $D_A=D_{A^\dag}$.
 A densely defined operator $B$ in a Hilbert space
 is said to be symmetric if
$\la\psi|B\vfi\ra=\la B\psi|\vfi\ra$ $\forall \psi,\vfi\in D_B$,
see e.g. \cite{HP}. In this paper we do not make difference
between symmetric and Hermitian operators.
}
 operator
$h$ related to
$H$ by an equivalence transformation (see e.g. \cite{myJPA_2011}).
In other
words, there is no way to redefine the inner product in the spirit of
paper \cite{Bender_Br_J} with respect to which $H$ would become Hermitian.

Probably, just for this reason some authors
claimed that for a Hamiltonian possessing the
spectral singularity no resolution of
 identity operator is possible \cite{collection,Most1}.
In paper \cite{ACS} the completeness of biorthogonal
sets of eigenfunctions of non-Hermitian
Hamiltonians possessing spectral singularities were carefully analyzed.
Obtained results are illustrated by a number of concrete examples.
In particular, the authors \cite{ACS} showed that
 the contribution of the spectral singularity to the resolution
 of identity operator
 depends on the class of functions employed for physical states.
 Further progress was made in \cite{myJPA_2010} where
  a special regularization procedure for the resolution of identity
  operator has been proposed.
We  also note the paper \cite{Gus} where a concise analysis of
 the general concept of the spectral singularity
 of  non-Hermitian  Hamiltonians is given.

Till now Hamiltonians possessing spectral singularities are studied
mainly as a possible source of new properties of optical media
\cite{collection}-\cite{Most2}.
Probably this is because of the fact that their association with
 quantum mechanical observables is involved.
 Nevertheless, as shown in \cite{ACS}, spectral singularities
 ``are physical'' since
 ``they contribute to transmission and reflection
coefficients of a non-Hermitian Hamiltonian
 dramatically enhancing their values''.

On the
other hand, any non-Hermitian diagonalizable Hamiltonian
$H$ with real and purely discrete  spectrum possesses a Hermitian
counterpart $h = h^\dag$ that is related to $H$ by a similarity
transformation \cite{Most_JMP2002}. Such a transformation does
not exist if the Hamiltonian is not diagonalizable.
(Such  Hamiltonian cannot be reduced to a diagonal form by changing the
basis. The interested reader can find a discussion about quantum mechanics
with non-diagonalizable Hamiltonians in \cite{Sokol}.)

For scattering Hamiltonians additional obstruction appears for
existing a similarity transformation between $H$ and $h$. Such a
transformation does not exist if in the continuous spectrum of
$H$ the spectral singularity is present.
In many cases $H$
depends on a parameter, $H=H(d)$ and the spectral
singularity appears at $d=d_0$.
If $d\ne d_0$ the
continuous spectrum of $H$ is regular.
If there exists an
invertible and positive definite operator $\eta$ such that
$\eta H(d)=H^\dag(d)\eta$, where $H^\dag(d)$ is Hermitian conjugate of
$H(d)$,
then one can construct Hermitian counterpart
$h(d)$ of the operator $H(d)$.
Below, using a very simple example,
we show that the possible presence
of a spectral singularity in $H(d)$,
for $d\approx d_0$
may be detected indirectly as a resonance in the
scattering cross section for $h(d)$.

In this paper we present a careful analysis of
Schwartz's example of a non-Hermitian Hamiltonian $H$ with the
possible presence of the spectral singularity.
This is one of the simplest examples since the Hamiltonian
contains only the kinetic energy and its non-Hermitian character is hidden
in a boundary condition at $x=0$.
In the next section we define this Hamiltonian and give a definition of
the spectral singularity. In Section \ref{sect3} we introduce a Hermitian
Hamiltonian $h$ together with $\eta$ operator which intertwines $H$ and
$H^\dag$. In the next section we reveal a supersymmetric nature of $\eta$
operator and introduce its superpartner $\wt\eta$.
In Section \ref{sect5} we construct an integro-differential operator which
being applied to the function $(2/\pi)^{1/2}\sin(kx)$ gives eigenfunctions
$\Phi_k(x)$ of $h$.
In the next section we calculate asymptotic behavior of $\Phi_k(x)$ and
scattering matrix for $h$. In Section \ref{sect7} we show that if $H$
has the spectral singularity, operator $h$ becomes undetermined.
In the last section we review shortly our main findings and
 draw some conclusions.

\section{Non-Hermitian Hamiltonian $H$\label{sect_H}}

Following Schwartz \cite{Schwartz} we
consider a non-Hermitian operator (Hamiltonian)
\be{Hd}
 H=-d^2/dx^2\,,\quad x\ge0
\ee
 with the domain
\be{DH}
D_H=\{\phi\in L^2(0,\infty):\ \phi''(x)\in L^2(0,\infty)\,,\
\phi'(0)+(d+ib)\vfi(0)=0\}
\ee
where $d$ and $b$ are real numbers.
This is a simple exercise to find its eigenfunctions $\phi_k(x)$
\[
H\phi_k=k^2\phi_k\,,\quad \phi_k\in D_H\,,
\]
\be{ex_vfik}
\vfi_k(x)=\sqrt{\frac{2}{\pi}}\,\bigl[\,k^2+(d+ib)^2\,\bigr]^{-1/2}
\left[
(d+ib) \sin(kx)-k\cos(kx)
\right]
\ee
which for $d<0$ form a bi-orthonormal set in $L^2(0,\infty)$
\be{}
\la\vfi^*_k|\vfi_{k'}\ra=\int_0^\infty \vfi_k(x)\vfi_k(x')\,dx
=\d(k-k')
\ee
with the completness condition of the form
\[
\int_0^\infty dk|\vfi_k\ra\la\vfi^*_k|=\bI\,.
\]
Here and in what follows we denote by $\bI$ identity operator and the
asterisk means the complex conjugation.

We would like to emphasize that if $d>0$
then the Hamiltonian $H$ has a discrete level
(see, e.g., \cite{Gus}), the possibility that we would like to avoid and
therefore in what follows we assume $d<0$.

There are several equivalent definitions of spectral singularities
(see e.g. \cite{Gus}). The one which is suitable for our purpose
uses the kernel $R(x,\xi,\lb)$ of the resolvent $R_\lb$ of $H$,
\[
R_\lb f(x)=\int_0^\infty R(x,\xi,\lb)f(\xi)\,d\xi\,.
\]
A point $\lb_0$ belonging to
 a continuous part of the spectrum of $H$ such that
 \[
 \lim_{\lb\to\lb_0} R(x,\xi,\lb)=\infty
 \]
where the limit should be taken along any path belonging to the resolvent
set of $H$.
The function $R(x,\xi,\lb)$ (see e.g. \cite{Gus}) is constructed with the
help of two linearly independent solutions $\vfi(x,\lb)$ and $e(\xi,k)$
of the differential equation
\be{vfipp}
-\vfi''(x)=\lb\vfi(x)\,,\quad\lb=k^2
\ee
 as follows
\be{resolvent}
\fl
R(x,\xi,\lb)=\frac{R_1(x,\xi,\lb)}{W(\lb)}\,,\quad
R_1(x,\xi,\lb)=\left\{\begin{array}{l}
\vfi(x,\lb)\,e(\xi,k)\quad \mbox{for} \ 0\le x\le\xi<\infty\,,
\\[.5em]
\vfi(\xi,\lb)\,e(x,k)\quad \mbox{for} \ 0\le \xi\le x<\infty\,.
\end{array}\right.
\ee
Here $W(\lb)$ is the Wronskian of the functions $\vfi(x,\lb)$
and $e(\xi,k)$,
\[
W(\lb)=\vfi(x,\lb)e'(x,k)-\vfi'(x,\lb)e(x,k)\,.
\]

In particular, if the function $\vfi(x,\lb)$ is such that
\[
\vfi(0,\lb)=1\,,\quad \vfi'(0,\lb)+(d+ib)\vfi(0,\lb)=0
\]
and $e(x,k)$ is the Jost solution of equation \rf{vfipp} defined by its
asymptotic behavior
\[
e(x,k)=e^{ikx}\,[1+o(1)]\,,\quad x\to\infty\,,\quad \mbox{Im}\,k\ge0
\]
then the Wronskian $W(\lb)$ coincides with the Jost function for the
Hamiltonian $H$
\[
W(\lb)=e'(0,k)+(d+ib)\,e(0,k)=ik+d+ib\,,\quad \lb=k^2\,.
\]
Since the resolvent becomes infinite at any point where $W(\lb)=0$,
i.e., in the current case
at $\lb=-(d+ib)^2$, the Hamiltonian $H$ has the spectral singularity
at point $\lb=b^2$, i.e., at $d=0$.
Just at this point, as it was mentioned in Introduction, the Jost function
for $H$ vanishes.
As shown in \cite{myJPA_2010} in this case the
corresponding continuous spectrum eigenfunction has zero binorm and the
resolution of identity operator needs a special regularization procedure.

\section{Equivalent Hermitian operator $h$ \label{sect3}}

To establish an equivalence between the non-Hermitian operator $H$ and
a Hermitian operator $h$,
we will use ideas formulated in \cite{SGH_92}
for quasi-Hermitian Hamiltonians and further developed
in \cite{Most_2003} for pseudo-Hermitian Hamiltonians. First one has to
find a Hermitian positive definite and invertible operator $\eta$ such
that
\be{etaH}
\eta H=H^\dag\eta\,.
\ee
In our case
the adjoint operator $H^\dag$ is defined by the same differential
expression \rf{Hd} with the domain
\be{DHD}
D_{H^\dag}=\{\phi\in L^2(0,\infty):\ \phi''(x)\in L^2(0,\infty)\,,
\ \phi'(0)+(d-ib)\vfi(0)=0\}.
\ee
It is not difficult to check that
a second order differential operator
\be{eta_oper}
\eta=-\p^2_x-2ib\p_x+d^2+b^2
\ee
 satisfies equation \rf{etaH}.
Evidently, equation \rf{etaH} defines $\eta$ up to
a transformation $\eta\to A^\dag\eta A$ with any invertible $A$
such that $[A,H]=0$ (see e.g. \cite{most2}).
We find convenient to use
the form \rf{eta_oper} of $\eta$ operator.

If $\eta$ were bounded its domain would be the whole Hilbert space
and no problems occurred to act by both the left and the right hand sides
of \rf{etaH} on functions belonging to $D_H$.
Unfortunately, this is not our case since
operator \rf{eta_oper} is
unbounded and should have
its own domain in $L^2(0,\infty)$.
 We find reasonable to assume that
 the domain of $\eta$ coincides with that of $H$,
\be{Deta}
D_\eta=D_H\,.
\ee
As we show below,
this assumption is justified by the property that
operator $\eta$  defined in this way is selfadjoint as well as
 positive definite and invertible on  $D_\eta$.
 It is not difficult to find its eigenfunctions and eigenvalues,
\be{etaPsik}
\eta\Psi_k(x)=(k^2+d^2)\Psi_k(x)\,,\quad k\ge0\,,
\ee
where
\be{Psik_x}
\Psi_k(x)=\sqrt{\frac{2}{\pi}}(k^2+d^2)^{-1/2}e^{-ibx}
\bigl[\,d\sin(kx)-k\cos(kx)\,\bigr].
\ee
From here we conclude that $\eta$ \rf{eta_oper} is positive
definite.
Moreover, since its spectrum is bounded below by $d^2\ne0$,
the operator $\eta^{-1}$ is bounded in $L^2(0,\infty)$ and can be
continued from any initial domain to the whole $L^2(0,\infty)$.

We note also that functions \rf{Psik_x} form a complete and orthonormal set in
$L^2(0,\infty)$,
\be{completPsik}
\la\Psi_k|\Psi_{k'}\ra=\d(k-k')\,,\quad
\int_0^\infty dk\,|\Psi_k\ra\la\Psi_k|=\bI\,.
\ee
This property follows form the fact that $\eta$ is seladjoint
with respect to the usual inner product in $L^2(0,\infty)$.
Indeed, as usual assuming that $\psi_1\in D_\eta$ and
integrating by parts twice  the term with the second derivative
and once the term with the first  derivative yields
\[
\fl
\begin{array}{lcl}
\la\psi_2|\eta\psi_1\ra
&=&  \displaystyle
\int_0^\infty\psi_2^*
[\,-\psi''_1+(w^*-w)\psi'_1+(ww^*-w')\psi_1\,dx\\[.7em]
&=&     \displaystyle
\left[\,\psi_2^*{}'\psi_1-\psi_2^*\psi'_1
+\psi_2^*\psi_1(w^*-w)\,\right]_{x=0}^\infty
\\[.7em]
& &+       \displaystyle
\int_0^\infty
[\,-\psi_2^*{}''+(w-w^*)\psi_2^*{}'\psi_1
-(w^*{}'-w')\psi_2^*\psi_1+(ww^*-w')\psi_1\psi_2^*\,]\,dx
\\[.7em]
&=&        \displaystyle
\int_0^\infty
[-\psi_2''+(w^*-w)\psi_2'+(ww^*-w')\psi_2\,]^*\psi_1\,dx
\\[.7em]
&=&\la\eta\psi_2|\psi_1\ra\,.
\end{array}
\]
To justify the last equality we consider the integrated term
at $x=0$
\ba{}\nn
\fl
\left[\,\psi_2^*{}'\psi_1-\psi_2^*\psi'_1
+\psi_2^*\psi_1(w^*-w)\,\right]_{x=0}
&=&
\left[\,\psi_2^*{}'\psi_1-\psi_2^*(-w\psi_1)
+\psi_2^*\psi_1(w^*-w)\,\right]_{x=0}
\\[.5em]
\nn
&=&
\psi_1(0)\left[\,\psi_2^*{}'+\psi_2^*w^*\,\right]_{x=0}
\\[.5em]
\nn
&=&0\,.
\ea
The first line here
 follows from the property that $\psi_1\in D_\eta=D_H$
given in \rf{DH} and in
the last line we used $\psi_2\in D_\eta=D_H$.

In
the next step we have to check that $\eta$ \rf{eta_oper} is
invertible on $D_\eta$.
For that we compute the kernel space
of the differential expression \rf{eta_oper}. This is a
two-dimensional linear space with the basis vectors
\[
f_\mp(x)=e^{-ibx\mp dx}\,.
\]
Evidently, $f_+(x)$ does not satisfy the boundary condition
given in equation \rf{DH} while $f_-(x)$ satisfies this
condition.
Therefore, for $d<0$, when
$f_-(x)\notin L^2(0,\infty)$, we have
$f_-(x)\notin D_\eta$ and, hence, $\eta$ \rf{eta_oper} is
invertible on $D_\eta$.
Thus,
as it was already mentioned,
in what follows we assume $d<0$ except for Section \ref{sect7}
where we consider the case $d=0$.

From Intertwining relation \rf{etaH} it follows that the operator defined
as
\be{h}
h=\eta^{1/2}H\eta^{-1/2}=\eta^{-1/2}H^\dag\eta^{1/2}
\ee
is Hermitian, $h=h^\dag$, and at the same time is related to $H$ by
equivalence transformation \rf{h}.
According to the first equality \rf{h} if $\psi_1\in D_h$ then
$\vfi_1=\eta^{-1/2}\psi_1$ should belong to $D_H$,
$\vfi_1\in D_H$.
Note that since both $\eta^{-1}$ and $\eta^{-1/2}$ are bounded,
the function
$\vfi_1$ is well defined.
Hence we can define $D^{(1)}_h$ as a set of
functions $\psi_1=\eta^{1/2}\vfi_1$ when $\vfi_1$ runs through
 $\wt D_H\subset D_H$ where $\wt D_H$ will be specified below.
 The function $\psi_1$ here is also well defined since
 $\wt D_H\subset D_H= D_\eta\subset D_{\eta^{1/2}}$.
 Similarly, according to the second equality in
\rf{h},
 we can define $D^{(2)}_h$ as a set of functions
$\psi_2=\eta^{-1/2}\vfi_2$ when $\vfi_2$ runs through
$\wt D_{H^\dag}\subset D_{H^\dag}$.
It is not difficult to see that for any $\vfi_1(x)$ satisfying the
boundary condition \rf{DH} the function
\be{fi12}
\vfi_2(x)=\eta\vfi_1(x)\in \wt D_{H^\dag}\subset D_{H^\dag}\,,\quad
(\vfi_1\in \wt D_H)
\ee
satisfies the boundary condition \rf{DHD}.
Note that since $\eta$ is a second order differential eexpresion
and $\vfi_2(x)\in D_{H^\dag}$,
the function $\vfi_1(x)$ should be
smoother than this is required by equation \rf{DH},
namely it
should be such that $\vfi^{(iv)}(x)\in L^2(0,\infty)$ where
$\vfi^{(iv)}(x)$ is the fourth derivative of $\vfi(x)$.
Thus we have
\[
\wt D_H=\{\vfi(x):\ \vfi(x)\in D_H\,,\ \vfi^{(iv)}(x)\in
L^2(0,\infty)\}\subset D_H\,.
\]
Moreover from \rf{fi12} it follows that
\[
\eta^{-1/2}\vfi_2(x)=\eta^{1/2}\vfi_1(x)\,.
\]
This means that we can put $D^{(1)}_h=D^{(2)}_h=D_h$ with
\[
D_h=\{
\psi(x):\ \psi(x)=\eta^{1/2}\vfi(x)\,,\
\vfi(x)\in \wt D_H\subset D_H
\}\,.
\]
Furthermore, since $\eta$ has an empty kernel on $D_H$, the set
$D_h$ is dense in $L^2(0,\infty)$ and it can be taken as an
initial domain for $h$ where it is Hermitian, i.e.
\[
\la\psi_2|h\psi_1\ra=\la h\psi_2|\psi_1\ra\quad\forall
\psi_1,\psi_2\in D_h\,.
\]
This property follows from the following chain of equalities:
\begin{eqnarray}\nn
\la\psi_2|h\psi_1\ra&=&\la\psi_2|h\eta^{1/2}\vfi_1\ra=
\la\psi_2|\eta^{1/2}H\vfi_1\ra
=\la\eta^{1/2}\psi_2|H\vfi_1\ra
\\[.5em] \nn
&=&\la\eta^{-1/2} H^\dag\eta^{1/2}\psi_2|\psi_1\ra=
\la h\psi_2|\psi_1\ra\,.
\end{eqnarray}

\section{SUSY partner of $\eta$ operator\label{sect4}}

As any positive definite second order differential operator,
$\eta$ admits a factorization
by first order operators $L$ and $L^\dag$,
\be{LLdag}
\eta=LL^\dag\,,\quad
L=-d/dx+d-ib\,,\ L^\dag=d/dx+d+ib
\ee
thus revealing its supersymmetric (SUSY) nature.
Corresponding SUSY algebra is based on the above factorization properties
and intertwining relations
(see e.g. \cite{myTMP})
\be{L_inter}
L^\dag\eta=\wt\eta L^\dag\,,\quad
\eta L=L\wt\eta
\ee
where
\be{LdagL}
 \wt\eta=L^\dag L\,.
\ee
Note that intertwining relations \rf{L_inter} are nothing but identities
\[
(L^\dag L)L^\dag=L^\dag (LL^\dag)\,,\quad
(LL^\dag) L=L(L^\dag L)\,.
\]
Although operator $\wt\eta$, which is a SUSY partner of $\eta$,
is defined by the same differential expression
as operator $\eta$ \rf{eta_oper},
its domain $D_{\wt\eta}$ is different than $D_\eta$ \rf{Deta}.
This, in particular, follows from intertwining relations \rf{L_inter}.
Indeed, according to these relations, operator $L^\dag$ transforms
eigenfunctions of $\eta$ to eigenfunctions of $\wt\eta$ and operator $L$
realizes an inverse mapping.
Taking unto account factorization properties \rf{LLdag} and \rf{LdagL},
we find
\be{Psimapping}
\wt\Psi_k=(k^2+d^2)^{-1/2}L^\dag\Psi_k\,,\quad
\Psi_k=(k^2+d^2)^{-1/2}L\wt\Psi_k\,.
\ee
Factor $(k^2+d^2)^{-1/2}$ guaranties the normalization of these
functions
\[
\la\Psi_k|\Psi_{k'}\ra=\d(k-k')\,,\quad
\la\wt\Psi_k|\wt\Psi_{k'}\ra=\d(k-k')\,.
\]
Thus using \rf{Psimapping} and \rf{Psik_x} one finds
the eigenfunctions of $\wt\eta$
\be{wtPsikx}
\wt\Psi_k(x)=\sqrt{\frac{2}{\pi}}\,e^{-ibx}\sin(kx)\,,
\quad\wt\eta\wt\Psi_k=(k^2+d^2)\wt\Psi_k\,.
\ee
Note that
these functions satisfy the Dirichlet boundary condition at $x=0$.
One can check that operator $\wt\eta$ defined on the domain
\[
D_{\wt\eta}=\{\Psi\in L^2(0,\infty):\ \Psi''(x)\in L^2(0,\infty)\,,
\quad \Psi(0)=0\}
\]
by the differential expression \rf{eta_oper}
is selfadjoint.

Evidently, the functions \rf{wtPsikx} are $d$-independent and
form an  orthonormal  and complete (in the sense of distributions)
 basis
in  $L^2(0,\infty)$
\be{}
\int_0^\infty dk\,|\wt\Psi_k\ra\la\wt\Psi_k|=\bI\,.
\ee

Another remarkable property of intertwining operators \rf{LLdag},
that we need below,
is the value of the composition
\be{h0a}
L^\dag L^*=-\frac{d^2}{dx^2}+(d+ib)^2\,,\quad
L^*=-d/dx+d+ib\,.
\ee

\section{Eigenfunctions of $h$\label{sect5}}

First we note that the eigenfunctions $\vfi_k$ of $H$ \rf{ex_vfik} may be
obtained by applying operator $L^*$ \rf{h0a} to the functions
\be{psik}
\psi_k(x)=\sqrt{\frac2\pi}\,\sin(k x)\,,
\ee
which yields
\[
\vfi_k(x)=[k^2+(d+ib)^2]^{-1/2}L^*\psi_k(x)\,.
\]
Therefore
the eigenfunctions $\Phi_k$ of $h$, obtained by operating with the
metric operator $\eta^{1/2}$ on the eigenfunctions of $H$,
may also be expressed in terms of the functions $\psi_k$
\begin{eqnarray}\nn \label{PhikL}
\Phi_k(x)&=&[k^2+(d+i b)^2]^{-1/2}\eta^{1/2}\vfi_k(x)
\\[.5em]
&=&[k^2+(d+ib)^2]^{-1}\eta^{1/2}L^*\psi_k(x)\,.
\label{PhikL}
\end{eqnarray}
Here the factor $[k^2+(d+ib)^2]^{-1/2}$ guaranties the
normalization of the functions on the Dirac-delta function.
We also note that $L^*\psi_k(x)\notin D_{\eta^{1/2}}$ but
this should not cause troubles since all continuous spectrum
eigenfunctions
are here generalized eigenfunctions of corresponding operators
and should be
understood in the sense of distributions.

Note that function \rf{psik} is an eigenfunction of the operator
\rf{h0a}
\be{LdLapsik}
L^\dag L^*\psi_k(x)=[k^2+(d+ib)^2]\psi_k(x)\,.
\ee

We find it useful expressing $\Phi_k(x)$ as result of the action on the
functions $\wt\Psi_k$ \rf{wtPsikx} by an
integro-differential operator.
To this end we first insert the identity operator
\rf{completPsik} between $\eta^{1/2}$ and $L$ in \rf{PhikL}
and use equation \rf{etaPsik}
\be{Phikp}
\Phi_k(x)=[k^2+(d+i b)^2]^{-1}
\int_0^\infty dk'\,\sqrt{(k'){}^2+b^2}\,\Psi_{k'}(x)
\la\Psi_{k'}|L^*\psi_k\ra
\ee
and then in the obtained expression replace $\Psi_k(x)$
according to \rf{Psimapping}
\be{Psikx}
\Phi_k(x)=L\int_0^\infty\frac{dk'}{\sqrt{(k')^2+d^2}}\,
\wt\Psi_{k'}(x)\la\wt\Psi_{k'}|\psi_k\ra\,.
\ee
Here we moved the operator $L$ from the left side in the inner product
to the right side where it becomes adjoint $L^\dag$,
replaced the action of the
the superposition $L^\dag L^*$ by its explicit expression \rf{h0a}
and used formula \rf{LdLapsik}.

An advantage of using the functions $\wt\Psi_k$ in \rf{Psikx}
with respect to using $\Psi_k$ in \rf{Phikp} is that the integral operator
in \rf{Psikx} is the positive and Hermitian square root of a resolvent
operator and therefore it is bounded in $L^2(0,\infty)$ whereas the
integral operator in \rf{Phikp} is unbounded.

\section{Asymptotic behavior of functions $\Phi_k(x)$.
Scattering matrix and cross section for $h$\label{sect6}}

Note that according to formula \rf{Psikx}, the eigenfunctions
$\Phi_k(x)$
of the
Hermitian operator $h$ are expressed in terms of elementary functions
\rf{wtPsikx} and \rf{psik}.
Nevertheless, no simple explicit expression for
these functions exists.
Below we calculate their asymptotic behavior as $x\to\infty$.

Inserting formulas \rf{h0a} and \rf{LdLapsik} into \rf{Psikx} yields
\be{PhikI}
\Phi_k(x)=
\left(\tfrac2\pi\right)^{3/2} L\,e^{-ibx}I(x)\,,\quad
I(x)=\int_0^\infty dy\,e^{iby}\sin(ky)\,
J(x,y)
\ee
where
\[
J(x,y)=\int_0^\infty dk'\,\frac{\sin(k'x)\sin(k'y)}%
{\sqrt{(k')^2+d^2}}\,.
\]
Expanding the product of sine functions into the difference of cosine
functions reduces the above integral to the one published in
\cite{GR} (see formula N 3.754.2),
\[
J(x,y)=\tfrac12\,K_0\bigl(d|y-x|\bigr)-\tfrac12\,K_0\bigl[d(y+x)\bigr]\,.
\]
Here $K_0(z)$ is the standard modified Bessel function
(see e.g. \cite{GR}).
Accordingly, integral from \rf{PhikI} has two contributions
\be{Ix}
I(x)=\tfrac12\,I_1(x)-\tfrac12\,I_2(x)
\ee
where the first term $I_1(x)$ contains the function
$K_0\bigl(d|y-x|\bigr)$
\[
I_1(x)=\int_0^\infty dy\,\sin(ky)\,K_0\bigl(d|y-x|\bigr)
\]
and the second term $I_2(x)$ is expressed in terms of the function
$K_0\bigl[d(y+x)\bigr]$
\[
I_2(x)=\int_0^\infty dy\,\sin(ky)\,K_0\bigl(d(y+x)\bigr)\,.
\]
With the change of the integration variable in the last
integral,
$d(y+x)=t$, and letting $x$ tend to infinity, we see that
\be{I20}
I_2(x)\to0\,.
\ee
Making a similar replacement
in the first integral,
$d(y-x)=t$, and also letting $x\to\infty$, we obtain a non-zero result
\[
I_1(x)\to\frac1d\,e^{ibx}I_{11}(x)
\]
where
\[
I_{11}(x)=\int_{-\infty}^\infty dt\,K_0\bigl(|t|\bigr)
\sin\bigl[k(x+t/d)\bigr]\,e^{ibt/d}\,.
\]
With the help of standard trigonometric formulas we reduce the product of
sine and exponential into a sum of four terms two of which are even and two
others are odd with respect to the replacement $t\to-t$.
 Because of the
symmetric integration limits, the odd terms vanish and
the integration limits in the integrals with the even terms
can be reduced to the semiaxis $(0,\infty)$. As a
result, both these terms reduce to the standard integral
(see \cite{GR}, equation No 6.671.14)
\[
\int_0^\infty K_0(x)\,\cos(\a x)\,dx=\frac{\pi}{2\sqrt{1+\a^2}}
\]
so that
\be{I11x}
I_{11}(x)=\frac{i\pi d}{2}\Biggl[\frac{e^{-ikx}}{\sqrt{d^2+(k-b)^2}}-
\frac{e^{ikx}}{\sqrt{d^2+(k+b)^2}}\Biggr]
\ee
Now using equations \rf{PhikI}, \rf{Ix}, \rf{I20} and \rf{I11x}
we finally obtain
\[
\Phi_k(x)\to\sqrt{\tfrac{2}{\pi}}\,e^{-i(\vfi_1-\vfi_2)/2}\,
\sin\bigl[kx+\tfrac12\,({\vfi_1+\vfi_2})\bigr]\,,\quad x\to\infty
\]
where
\[
\vfi_1=\frac{1}{2i}\,\log\Bigl[\frac{d-ik+ib}{d+ik-ib}\Bigr]\,,\quad
\vfi_2=\frac{1}{2i}\,\log\Bigl[\frac{d-ik-ib}{d+ik+ib}\Bigr]\,.
\]
From here we find the phase shift
\be{phase-shift}
\d=\tfrac12\,(\vfi_1+\vfi_2)=\frac{1}{4i}\,
\log\Bigl[\frac{b^2+(d-ik)^2}{b^2+(d+ik)^2}\Bigr]
\ee
and the $S$-matrix
\be{S-matr}
S=e^{2i\d}=\Bigl[\frac{b^2+(d-ik)^2}{b^2+(d+ik)^2}\Bigr]^{1/2}\,.
\ee
This result perfectly agrees with the general formula for the $S$-matrix
obtained in \cite{myJPA_2011}.

We note that the scattering matrix
\be{S2R}
S_{BW}=S^{\,2}=\frac{b^2+(d-ik)^2}{b^2+(d+ik)^2}
\ee
leads to
a Breit-Wigner resonance formula (see e.g. \cite{Perkins})
\[
\sigma_{BW}=\frac{16\pi d^2}{(k^2+d^2-b^2)^2+4b^2d^2}
\]
which in the energy scale  reads
\be{sgBW}
\sigma_{BW}=\frac{4 \pi}{b^2}\,\frac{(\G/2)^2}{(E-E_0)^2+(\G/2)^2}
\ee
with $\G=4 bd$ and $E_0=b^2-d^2$.
We assume that $|d|$ is small enough so that $b^2>d^2$.
Near the resonance $E\approx E_0$ and $k\approx b$ so that Eq. \rf{sgBW}
 reduces to the celebrated Breit-Wigner formula
 (see e.g. \cite{Bohm}).
 From here we conclude that the $S$-matrix \rf{S-matr} is a square root of the
 Breit-Wigner $S$-matrix $S_{BW}$ given in \rf{S2R}.

 The phase shift $\d$ \rf{phase-shift}
  corresponding to $S$ \rf{S-matr} is one half of $\d_{BW}$,
$\d_R=\frac12\d_{BW}$.
It leads to a cross section with a square root branch point
\cite{myJPA_2011}
\be{sgR}
\sg(k)=\frac{2\pi}{k^2}\biggl[
1+\frac{k^2-b^2-d^2}{\sqrt{(k^2+d^2-b^2)^2+4b^2d^2}}
\biggr].
\ee
We choose here that sign of the square root
which corresponds to positive definite operator $\rho=\eta^{1/2}$
\cite{myJPA_2011}.
It is not difficult to see that
$\sg(0)=4\pi d^2/(b^2+d^2)^2>0$, $\lim_{k\to\infty}\sg_R(k)=0$,
$\sg(k)>0$ for $0\le k<\infty$ and $d\sg_R(0)/dk>0$ for
$b^2>\frac12d^2$.
These results mean that for any fixed value of $b$ and small enough value
of $|d|$, the function $\sg(k)$ \rf{sgR} has a maximum and therefore exhibits
a resonance behavior.
This  is just a consequence of the fact that
the Hamiltonian $H$ is, in a sense, close to that which has
 a spectral singularity.

\section{Spectral singularity, $d=0$\label{sect7}}

As was discussed in Section \ref{sect_H}
spectral singularity appears in $H$
only when $d=0$.
Note that the functions $\Psi_k$ \rf{Psik_x}
 as well as operator $\eta$ has no
singularity at $d=0$:
\be{}
\Psi_k(x)=\sqrt{\frac2\pi}\,\,e^{-ibx}\cos(kx)\,.
\ee
Thus operator $\eta$ is well defined in $L^2(0,\infty)$ at
the spectral singularity of $H$ and
its eigenfunctions form an orthonormal and complete
set in $L^2(0,\infty)$,
\be{}
\la\Psi_k|\Psi_{k'}\ra=\d(k-k')\,,\quad
\int_0^\infty dk\,|\Psi_k\ra\la\Psi_k|=1\,.
\ee
It ia apparent that operator $\rho$, being positive and
Hermitian square root of $\eta$
\be{}
\rho=\eta^{1/2}=\int_0^\infty dk\,\sqrt{k^2+d^2}\,|\Psi_k\ra\la\Psi_k|
\ee
is also well defined for $d=0$.
This means that using this operator one is able to construct a physical
Hilbert but this does not mean that the non-Hermitian Hamiltonian $H$
 will be
mapped to a Hermitian Hamiltonian by a similarity transformation.
To illustrate this impossibility we will calculate eigenfunctions $\Phi_k(x)$
of $h$ at $d=0$.

Let us denote
\be{Axy}
A(x,y)=\int_0^\infty \frac{dk'}{k'}\wt\Psi_{k'}(x){\wt\Psi_{k'}}^*(y)\,.
\ee
Then the integro-differential operator \rf{Psikx}, applied to function
$\psi_k(y)$, yields
\be{APhik}
\Phi_k(x)=L\int_0^\infty A(x,y)\,\psi_k(y)\,dy\,.
\ee
Integral in \rf{Axy} with functions $\wt\Psi_k(x)$ given in \rf{wtPsikx}
is standard (see \cite{GR}, formula No 3.741.1) so that
the kernel $A(x,y)$ reads
\be{Aform}
A(x,y)=\frac1\pi\,e^{ib(y-x)}\log\left|\frac{x+y}{x-y}\right|\,.
\ee
Further integration in \rf{APhik} with $\psi_k$ given in \rf{psik}
can also be made explicitly if one uses formulas 4.382.1 and 4.382.2 from
\cite{GR}. Finally after some tedious calculations
assuming, for instance, $b>0$,
one gets

\[\fl
\begin{array}{lll}
\Phi_k(x)&=&\sqrt{\frac{2}{\pi^3}}\,\,e^{-ibx}\\[.5em]
& &\times i\bigl\{\,
\frac12\cos[(b-k)x]\,(Ci[(b-k)x]+Ci[(k-b)x])-
\cos[(b+k)x]\ci[(b+k)x]
\\[.5em]
& &+\sin[(b-k)x]\,\si[(b-k)x]
-\sin[(b+k)x]\,\si[(b+k)x]\,\bigr\}\\[.5em]
& & +\sqrt{\frac{2}{\pi}}\,\,e^{-ibx}\sin(bx)\sin(kx)
+\frac1{\sqrt{2\pi}}\,\frac{|b-k|}{b-k}\,e^{-ibx}\cos[(b-k)x]
\end{array}
\]
were
\[
\ci(z)=-\int_z^\infty\frac{\cos z}{z}\,dz\,,\quad
\si(z)=\int_0^z\frac{\sin z}{z}\,dz\,.
\]
We note that the last term here is undetermined for $k=b$.
We thus conclude that the point $k=b$ cannot belong to the continuous
spectrum neither it can belong to a discrete spectrum.
Therefore an operator that
has such eigenfunctions cannot be Hermitian in $L^2(0,\infty)$.
This conclusion is also supported by the fact that for $k=b$ the imaginary
part of the integral in \rf{APhik} is divergent.

This result is not surprising. Indeed, at $d=0$ the
eigenfunction of $H$ corresponding to $k=\pm b$ is proportional
to $\exp(-ibx)\in {\rm ker}\,\eta$. Although $\eta$ remains
invertible on $D_H$, it is not invertible when applied to
generalized eigenfunctions of $H$.

\section{Conclusion}

In this paper we analyzed one of the simplest non-Hermitian Hamiltonian $H$,
which at a specific value of a parameter
may possesses a spectral singularity in its continuous spectrum,
first proposed by Schwartz \cite{Schwartz}. It contains only kinetic
energy but the functions from its domain of definition satisfy a complex
boundary condition at $x=0$.
We have shown that $\eta$ operator
($\eta=\rho^2$) is a second order
differential operator with constant coefficients and revealed its
supersymmetric nature.
This approach permitted us to express
eigenfunctions $\Phi_k(x)$ of $h$, where $h$ is Hermitian and related to $H$ by a
similarity transformations, in terms of a bounded integral operator defined in
the Hilbert space $L^2(0,\infty)$.
With the help of this bounded operator
we succeeded to find asymptotic behavior of the functions $\Phi_k(x)$ and
calculate the scattering matrix and cross section for $h$.
Finally we have shown that at the
point in the parameter space
where $H$ has the spectral singularity the Hermitian operator $h$
becomes undetermined.
 Thus, using this specific
 example we demonstrated that a non-Hermitian
Hamiltonian possessing the spectral singularity cannot be mapped to a
Hermitian Hamiltonian by any similarity transformation.
Nevertheless, the possible presence of the spectral singularity in $H$ may
be detected as a resonance in the scattering cross section in $h$.

\end{document}